\documentclass[twocolumn,aps,prb,showpacs,amsmath,superscriptaddress]{revtex4}
        \usepackage{amsmath}
        \usepackage{amssymb}
        \usepackage{subfigure}
        \usepackage{natbib}
        \usepackage{epsfig}
        \usepackage{amsfonts}
        \usepackage{mathrsfs}
        \usepackage{ulem}
        \usepackage{color}
        \usepackage[toc,page,title,titletoc,header]{appendix}
        \usepackage{CJK}
        \usepackage{epstopdf}
    \usepackage{mathrsfs}
    \usepackage{float}
    \usepackage{tikz-cd}
    \usepackage{caption}
    \usepackage{graphicx, subfig}
    \usepackage{dcolumn}
    \usepackage{bm}
    \usepackage{float}

\begin{document}
      \widetext
      \title{Topological classification of non-Hermitian systems with reflection symmetry}
         \author{Chun-Hui Liu}
\affiliation{Beijing National Laboratory for Condensed Matter Physics, Institute of Physics, Chinese Academy of Sciences, Beijing 100190, China}
\affiliation{School of Physical Sciences, University of Chinese Academy of Sciences, Beijing 100049, China}
\author{Hui Jiang}
\affiliation{Beijing National Laboratory for Condensed Matter Physics, Institute of Physics, Chinese Academy of Sciences, Beijing 100190, China}
\affiliation{School of Physical Sciences, University of Chinese Academy of Sciences, Beijing 100049, China}
\author{Shu Chen}
\email{schen@iphy.ac.cn}
\affiliation{Beijing National Laboratory for Condensed Matter Physics, Institute of Physics, Chinese Academy of Sciences, Beijing 100190, China}
\affiliation{School of Physical Sciences, University of Chinese Academy of Sciences, Beijing 100049, China}
\affiliation{The Yangtze River Delta Physics Research Center, Liyang, Jiangsu 213300, China}
          \begin{abstract}
           \par
           We classify topological phases of non-Hermitian systems in the Altland-Zirnbauer classes with an additional reflection symmetry in all dimensions.
           By mapping the non-Hermitian system into an enlarged Hermitian Hamiltonian with an enforced chiral symmetry, our topological classification is thus equivalent to classifying Hermitian systems with both chiral and reflection symmetries, which effectively change the classifying space and shift the periodical table of topological phases.
           According to our classification tables, we provide concrete examples for all topologically nontrivial non-Hermitian classes in one dimension and also give explicitly the topological invariant for each nontrivial example. Our results show that there exist two kinds of topological invariants composed of either winding numbers or $\mathbb{Z}_2$ numbers. By studying the corresponding lattice models under the open boundary condition, we unveil the existence of bulk-edge correspondence for the one-dimensional topological non-Hermitian systems characterized by winding numbers,  however we did not observe the bulk-edge correspondence for the $\mathbb{Z}_2$ topological number in our studied $\mathbb{Z}_2$-type model.
          \end{abstract}
          \maketitle
          \section{Introduction}
          The band theory of topological insulators and superconductors has been greatly
          developed in the past decades \cite{Hassan,Qi,Wen1,Thouless,Haldane,Kane1,Kane2,Fu1,Fu2,ZhangSC}. Topologically non-trivial insulators
          or superconductors are characterized by either an integer ($\mathbb{Z}$) or a binary ($\mathbb{Z}_2$) topological index and have stable edge states \cite{Bernevig,Moore,Kitaev1}. The well-known examples are
           the integer quantum Hall effect and spin Hall effect, for which different
           topological phases are characterized by Chern number \cite{Thouless,Haldane} and $\mathbb{Z}_2$ topological number \cite{Kane1,Kane2,ZhangSC}, respectively. Traditionally, non-interacting fermionic systems can be divided into ten Altland-Zirnbauer (AZ) symmetry classes, in terms of the presence or absence of Time-reversal (T), charge-conjugation or particle-hole (C), and chiral or sublattice symmetry (S). In the framework of the 'ten-fold way', it is well established that five of ten classes are topologically nontrivial in every spatial dimension  \cite{Fu1,Fu2,AZ,Kitaev2,Ludwig1,
            Ludwig2,Ludwig3,Ludwig4}.  When adding additional symmetries to the system, the classifying space of Hamiltonian
            is changed, and we can get different topological classification \cite{Fu1,Fu2,AZ,Kitaev2,Ludwig1,
            Ludwig2,Ludwig3,Ludwig4,Furusaki,Stone,Chiu,Wen2}.  Many theoretical works have been carried out on
            the classification of various symmetry-protected-topological phases \cite{Furusaki,Stone,Chiu,Wen2}, ranging from the reflection symmetry protected topological phases to the crystalline topological phases, and most of
            these works are focused on the Hermitian systems.

            Recent experimental progress on the study of optical systems and electrical systems with gain and loss has unveiled that these systems can be effectively described by non-Hermitian Hamiltonians \cite{Ruter2010,Peng2014,Feng2014,Konotop2016,Xiao2017,Weimann2017,Menke,Klett,Bender1,Bender2,AUeda,Longhi}. The effect of gain and loss on topological properties of periodical lattice systems was studied within the non-Hermitian lattice models with parity-time symmetry \cite{Hu2011,Esaki2011,Rudner,ZhuBG,Yuce}. Except from the inherent gain or loss, non-Hermiticity can also arise from non-reciprocity \cite{Hatano}.  Recently, there is increasing interest in exploring topological phases in non-Hermitian systems \cite{Fu3,Fu4,Xiong,Xiao,Weimann,ElG,TElee1,TElee2,Leykam,Parto,Gong,Torres2,Bergholtz,WangZhong1,Torres1}. Various topological non-Hermitian models have been studied \cite{Yin,JiangHui,TElee2,Leykam,Lieu,Shen,WangZhong2,Song},
            including non-Hermitian extensions of Su-Schrieffer-Heeger (SSH) model \cite{Yin,JiangHui,Lieu}, Kitaev chain \cite{Song} , and two-dimensional Chern insulators \cite{Shen,WangZhong2}. It has been demonstrated that topological non-Hermitian systems may exhibit quite different behaviors from  their Hermitian counterparts, associated with some distinctive properties of the non-Hermitian Hamiltonian, e.g., the existence of exceptional points \cite{Heiss,Dembowski,Rotter,Hu2017,Hassan2017,Kim}, biorthonormal eigenvectors \cite{Bergholtz}, unusual bulk-edge correspondence \cite{Xiong,Bergholtz,WangZhong1,Torres1} and emergence of non-Hermitian skin effect  \cite{Xiong,Torres1,WangZhong1}.

            For systematic understanding of non-Hermitian topological phases, it is highly desirable to carry out topological classification for non-Hermitian systems.  Regardless of spatial symmetry, non-Hermitian systems can be classified into 43 categories \cite{Bernard}, which is more complicate than the Hermitian systems. Very recently, Gong et. al. have classified non-Hermitian phases with non-spatial symmetries \cite{Gong} for
             10 of the 43 categories by considering T, C and S symmetries, which can be viewed as an extension of topological classification of the non-Hermitian counterparts of AZ classes. Inspired by the periodic classification table for Hermitian topological phases protected by the reflection symmetry \cite{Furusaki,Chiu}, in this work we study the topological classification of non-Hermitian systems with reflection symmetry.
             So far, the classification of non-Hermitian systems is only limited to systems with internal (nonspatial) symmetries \cite{Gong}, and the classification for non-Hermitian systems with spatial symmetries is still lack. As the introduction of spatial symmetries leads to rich symmetry-protected topological phases, it is expected that the topological classification by considering the spatial symmetries is also important for non-Hermitian systems. As shown in this work, by considering the reflection symmetry, the non-Hermitian systems indeed exhibit rich reflection-symmetry protected topological phases in all dimensions, in contrast to the classification with only internal symmetries \cite{Gong}, according to which the two-dimensional non-Hermitian systems are always trivial.

             Our classification is based on K theory and Clifford algebras \cite{Kitaev2} and includes two steps. At the first step,  we smoothly deform the non-Hermitian Hamiltonian to a unitary matrix, and then map the unitary matrix to an enlarged band-flattened Hermitian matrix with an enforced chiral symmetry. This step is called Hermitianization \cite{RH,Gong}.
             The second step is to represent the classifying space as Clifford algebra extension. After the two steps, we get a richer classification with periodic classification tables for the complex and real classes shown in table II and III, respectively. Then we construct the topological invariant
             for each of the one-dimensional (1D) non-trivial classes and discuss the bulk-edge correspondence for
             various models.

              The paper is organized as follows. In section II, we first give a general description of classification of non-Hermitian system in the AZ classes with additional reflection symmetry and then give the classification tables. In section III, focusing on the one-dimensional non-Hermitian systems, we consider all topologically nontrivial examples according to our classification tables and construct topological invariants for all example Hamiltonians. In Section IV, we study the bulk-edge correspondence by considering  several typical example systems. A summary is given in the last section.

    \section{Classification of non-Hermitian system with reflection symmetry}
    The topological classification of non-Hermitian systems with time reversal symmetry $T$,
    pseudo particle-hole symmetry $C$ and chiral symmetry $S=TC$ was recently studied by Gong et.al. \cite{Gong}
    For a given AZ class which fulfills $T$ or $C$ symmetry, its Hamiltonian $H$ fulfills
    \begin{equation}
        AH(-k)=\eta _A H(k)A ,
    \end{equation}
where $\eta _A=\pm 1$ and $A$ is an anti-unitary operator with $A^2=\epsilon_A=\pm1$. We can always represent the anti-unitary operator $A$ as $A=U_A K$ with $U_A$ being an unitary matrix and $K$ being a complex conjugate operator. The operator $A$ can represent $T$ and $C$ with $\eta _T=1$ and $\eta _C=-1 $, respectively. For a Hamiltonian with the constraint $det(H(k)) \not=0 $, corresponding to $H(k)$ having no zero eigen energy,  one can always make a polar decomposition
\begin{equation}
H(k)=U(k)P(k),
\end{equation}
where $H(k)$ is a reversible matrix, $P(k)$ is a positive-definite Hermitian matrix and $U(k)$ is a unitary matrix. It was proved that $H(k)$ can be continuously deformed to $U(k)$ under $T$, $C$ or $S$ symmetry without the change of topological properties \cite{Gong}. So classifying $H(k)$ equals to classifying $U(k)$.  Here the continuous deformation agrees with the definition in Ref.\cite{Gong}. Given that the Hamiltonian $H_{\lambda}(k)$ ($0\leq \lambda \leq 1$) is a continuous function of $\lambda$ and $k$, if $det(H_\lambda(k)) \neq 0$  for all $\lambda \in [0,1]$, 
then $H_\lambda(k)$ is a continuous deformation. If the complex bands of a non-Hermitian Hamiltonian never cross at a point in the complex-energy plane, we can set this base point as the zero point by making a total shift of the spectrum in analogy with the Hermitian case where the Fermi energy $E_F$ is usually set to be zero.

To classify the non-Hermitian Hamiltonian, it is convenient to introduce an enlarged Hamiltonian
             \begin{equation}
                H_a(k)=\left[ \begin{array}{cc}
                    0 & U(k)\\
                    U(k)^\dagger & 0
                    \end{array}
                    \right ]. \label{Ha}
                \end{equation}
It is obvious that $H_a(k)$ is a Hermitian matrix, which fulfills a chiral symmetry
                    \begin{equation}
                        \Sigma H_a(k)=  -H_a(k) \Sigma , \label{chiralsym}
                        \end{equation}
where $\Sigma = \sigma_z \otimes 1$ and $\Sigma^2=1$.
Now the problem transforms to classify a Hermitian Hamiltonian $H_a(k)$ with an additional symmetry $\Sigma$. Under such a scheme, the classification has been given in Ref. \cite{Gong}. For convenience, we also list the results here (table I).

    \begin{table*}[htbp]
        \caption{\label{tab:table1}Classification of non-Hermitian systems with time reversal symmetry $T$, pseudo particle-hole symmetry $C$ and chiral symmetry $S=TC$.}
      \begin{center}
      \begin{tabular}{cccccccccc}
       \hline\hline
       \;\;\; Cartan class\;\;\;&\;\;\;Classifying Space\;\;\;&\;\;\;\;$d=0$\;\;\;\;&\;\;\;\;\;\;1\;\;\;\;\;\;&\;\;\;\;\;\;2\;\;\;\;\;\;&\;\;\;\;\;\;3\;\;\;\;\;\;&\;\;\;\;\;\;4\;\;\;\;\;\;&\;\;\;\;\;\;5\;\;\;\;\;\;&\;\;\;\;\;\;6\;\;\;\;\;\;&\;\;\;\;\;\;7\;\;\;\;\;\;\\
        \hline\hline
        Complex case\\
        A&$C_1$&$0$& $\mathbb{Z}$&0&$\mathbb{Z}$&0&$\mathbb{Z}$&0&$\mathbb{Z}$\\
        AIII&$C_1 \times C_1$&$0$&$ \mathbb{Z} \oplus \mathbb{Z}$&0&$\mathbb{Z} \oplus \mathbb{Z}$&0&$\mathbb{Z} \oplus \mathbb{Z}$&0&$\mathbb{Z} \oplus \mathbb{Z}$\\
        Real case\\
        DIII,CI&$C_1$&$0$& $\mathbb{Z}$&0&$\mathbb{Z}$&0&$\mathbb{Z}$&0&$\mathbb{Z}$\\
        AI,D&$R_1$&$\mathbb{Z}_2$&$ \mathbb{Z}$&0&$0$&0&$2\mathbb{Z}$&0&$\mathbb{Z}_2$\\
        AII,C&$R_5$&$0$&$ 2\mathbb{Z}$&0&$\mathbb{Z}_2$&$\mathbb{Z}_2$&$\mathbb{Z}$&0&$0$\\
        BDI&$R_1 \times R_1$&$\mathbb{Z}_2 \oplus \mathbb{Z}_2$&$ \mathbb{Z}\oplus \mathbb{Z}$&0&$0$&0&$2\mathbb{Z}\oplus 2\mathbb{Z}$&0&$\mathbb{Z}_2\oplus \mathbb{Z}_2$\\
        CII&$R_5 \times R_5$&$0$&$ 2\mathbb{Z} \oplus 2\mathbb{Z}$&0&$\mathbb{Z}_2 \oplus \mathbb{Z}_2$&$\mathbb{Z}_2 \oplus \mathbb{Z}_2$&$\mathbb{Z} \oplus \mathbb{Z}$&0&$0$\\
      \hline\hline
        \end{tabular}
      \end{center}
    \end{table*}

When the system has a reflection symmetry, we have
\begin{equation}
    R_lH(k_1,...,-k_l,...,k_d)=  H(k_1,...,k_l,...,k_d)R_l,
\end{equation}
where the reflection operator fulfills $R_l^2=1$.
By using the polar decomposition $H(k)= U(k)P(k)$, we can prove that $H(k)$ can be continuously deformed to $U(k)$ under the reflection symmetry (see the appendix A), and thus classifying H(k) is equivalent to classifying U(k). Similarly, we can define an enlarged Hermitian Hamiltonian $H_a$ given by Eq.(\ref{Ha}).
Besides the chiral symmetry given by Eq.(\ref{chiralsym}), $H_a$ also fulfills  the following symmetries:
              \begin{equation}
                  A_1 H_a(-k)=\eta _A H_a(k) A_1
                  \end{equation}
                  and
                            \begin{equation}
                                RH_a(k_1,...,-k_l,...,k_n)=  H_a(k_1,...,k_l,...,k_n)R, \label{reflection}
                                \end{equation}
where
\begin{equation*}
          A_1= \sigma_0 \otimes A
          \end{equation*}
and
              \begin{equation*}
                R=\sigma_0 \otimes R_l .
                \end{equation*}
Here $A$ can represent $T$ or $C$, i.e., $T_1= \sigma_0 \otimes T$ and  $C_1= \sigma_0 \otimes C$.
              Classifying U(k) equals to classifying Hermitian Hamiltonian $H_a(k)$
               with two addition symmetry $\Sigma$ and $R$. Since $H_a(k)$ is Hermitian,
               we transform the problem into a classification problem of Hermitian Hamiltonian.

                 Following the topological classification with additional symmetries from
                 Clifford algebras \cite{Furusaki,Chiu}, we can represent the Hamiltonian as
                 \begin{equation}
                    H_a(k)=\gamma_0+\sum_{i=1}^d\gamma_i k_i ,
                    \end{equation}
where $k_i$ denotes the momentum in the $i$th direction and the gamma matrices $\gamma_i$ satisfy the Clifford algebra, i.e.,
                    \begin{equation}
                        \left\{\gamma_i,\gamma_j\right\}=2\delta_{ij}.
                        \end{equation}
For $j\not=l$, we have $[R,\gamma_j]=0$. When $j=l$, $\left\{R,\gamma_l\right\}=0$ according to Eq.(\ref{reflection}). We also have $[R,\Sigma]=0$, $R^2=1$ and $\left\{\Sigma,\gamma_i\right\}=0$ (i=0,1,...,d) according to Eq.(\ref{chiralsym}). Define
                    \begin{equation}
                        M=JR\gamma _l,
                        \end{equation}
where $J=i$ is a complex structure. It follows that $M$ and $\Sigma$ satisfy
                    \begin{equation*}
                        \left\{\Sigma,M \right\}=0, \qquad
                        M^2=1, \qquad
                        \Sigma^2=1,
                        \end{equation*}
                          and
                          \begin{equation*}
                                \left\{M,\gamma_i\right\}=0 .
                                \end{equation*}
                                   \begin{table*}[htbp]
                                        \caption{\label{tab:table3} Classification of non-Hermitian systems for complex classes in the AZ symmetry classes in the presence of reflection symmetry. Consider chiral symmetry
                                        S and reflection symmetry $R_l$. For $R_lS=\eta _SSR_l$, we denote the commutation relation as $R^{\eta_S}$. }
                                      \begin{center}
                                      \begin{tabular}{cccccccccc}
                                      \hline\hline
                                      \;\;\; Cartan class\;\;\;&\;\;\;Classifying Space\;\;\;&\;\;\;\;$d=0$\;\;\;\;&\;\;\;\;\;\;1\;\;\;\;\;\;&\;\;\;\;\;\;2\;\;\;\;\;\;&\;\;\;\;\;\;3\;\;\;\;\;\;&\;\;\;\;\;\;4\;\;\;\;\;\;&\;\;\;\;\;\;5\;\;\;\;\;\;&\;\;\;\;\;\;6\;\;\;\;\;\;&\;\;\;\;\;\;7\;\;\;\;\;\;\\
                                        \hline
                                        A&$C_0$&$\mathbb{Z}$& 0&$\mathbb{Z}$&0&$\mathbb{Z}$&0&$\mathbb{Z}$&0\\
                                        AIII\qquad$R^+$&$C_0$&$\mathbb{Z}$& 0&$\mathbb{Z}$&0&$\mathbb{Z}$&0&$\mathbb{Z}$&0\\
                                        AIII\qquad$R^-$&$C_1$&$0$& $\mathbb{Z}$&0&$\mathbb{Z}$&0&$\mathbb{Z}$&0&$\mathbb{Z}$\\
                                       \hline\hline
                                        \end{tabular}
                                      \end{center}
                                    \end{table*}

                                 Topological classification of AZ classes with additional symmetries
                                has been studied in \cite{Furusaki,Chiu}.
                                Considering a Hermitian Hamiltonian, if we add multiple additional symmetries $\{ M_i \}$ to the classification ($\{M_i,H\}=0$), and the additional symmetries anticommute with each other $\{ M_i,M_j \}=2\delta _{i,j}$, we have the following conclusions by Clifford algebras and their extensions \cite{Furusaki}: For Class A, if the number of additional symmetries $M_i$ is $n$, the classifying space $C_0$ shifts to $C_{n}$; For Class AIII, if additional symmetries $M_i$ and chiral symmetry $S$ have the following relations: $M_iS=-SM_i$ for $1\leq i \leq m$ and $M_iS=SM_i$ for $m+1\leq i \leq n-m$,
                                the classifying space $C_1$ shifts to $C_{m+1}$.  With the help of the above conclusions,  we can get the non-Hermitian classification with reflection symmetry for the complex classes as shown in table II:

                                {\bf Class A:} Two symmetries $M$ and $\Sigma$ are added to the system ($n=2$), and thus the classifying space $C_0$ shifts to $C_2\simeq C_0$. Consequently, the non-Hermitian system of Class A with reflection symmetry is characterized by $\pi_d(C_0) = \mathbb{Z}$ $(0)$ for even (odd) d.

                                {\bf Class AIII($R^+$)} and {\bf  AIII($R^-$):} Here $R^{\eta_S}$ represents that $R_l S=\eta _S SR_l$ with $\eta_S = \pm 1$.
                                Adding two symmetries $M$ and $\Sigma$ to the system of the $H_a$, while $\Sigma=\sigma_z\otimes 1$ always commutes with $S_1=T_1 C_1 = \sigma_0 \otimes S $,  $M=JR\gamma_l$ anti-commutes with $\Sigma$ for Class AIII($R^+$) or commutes with $\Sigma$ for Class AIII($R^-$). 
                                It follows that $m=1$ for the Class AIII($R^+$), and the classifying space $C_1$ shifts to $C_2\simeq C_0$, suggesting that the system is characterized by $\pi_d(C_0) = \mathbb{Z}$ $(0)$ for even (odd) d. On the other hand, we have $m=0$ for the Class AIII($R^-$), and thus the classifying space $C_1$ keeps invariant, suggesting that the system is characterized by $\pi_d(C_1) = 0$ $(\mathbb{Z})$ for even (odd) d.

                              \begin{table*}[htbp]
                                    \caption{\label{tab:table4} Classification of non-Hermitian systems for real classes in the AZ symmetry classes in the presence of reflection symmetry. We consider time reversal symmetry T, pseudo particle-hole symmetry
                                    C and reflection symmetry $R_l$. Given that $R_lT=\eta _TTR_l$ and $R_lC=\eta _CCR_l$, if there
                                    is T (C) symmetry, we denote the commutation relations as $R^{\eta _T (\eta _C)}$;
                                    if there are both T and C symmetries, we denote the commutation relations as $R^{\eta _T \eta _C}$ .}
                                  \begin{center}
                                  \begin{tabular}{cccccccccc}
                                      \hline\hline
                                      \;\;\; Cartan class\;\;\;&\;\;\;Classifying Space\;\;\;&\;\;\;\;$d=0$\;\;\;\;&\;\;\;\;\;\;1\;\;\;\;\;\;&\;\;\;\;\;\;2\;\;\;\;\;\;&\;\;\;\;\;\;3\;\;\;\;\;\;&\;\;\;\;\;\;4\;\;\;\;\;\;&\;\;\;\;\;\;5\;\;\;\;\;\;&\;\;\;\;\;\;6\;\;\;\;\;\;&\;\;\;\;\;\;7\;\;\;\;\;\;\\
                                    \hline
                                    AI,D \qquad $R^+$&$R_2$&$\mathbb{Z}_2$&$\mathbb{Z}_2 $&$\mathbb{Z}$&0&0&0&$2\mathbb{Z}$&0\\
                                  AII,C\qquad$R^+$&$R_6$&$0$& 0&$2\mathbb{Z}$&0&$\mathbb{Z}_2$&$\mathbb{Z}_2$&$\mathbb{Z}$&0\\
                                  AI,D\qquad$R^-$&$R_0$& $\mathbb{Z}$&0&0&0&$2\mathbb{Z}$&0&$\mathbb{Z}_2$&$\mathbb{Z}_2$\\
                                  AII,C\qquad$R^-$&$R_4$&$2\mathbb{Z}$&0&$\mathbb{Z}_2$&$\mathbb{Z}_2$&$\mathbb{Z}$&0&0&0\\
                                    DIII,CI \qquad $R^{++}$&$C_0$&$\mathbb{Z}$& 0&$\mathbb{Z}$&0&$\mathbb{Z}$&0&$\mathbb{Z}$&0\\
                                    BDI\qquad$R^{++}$&$R_2 \times R_2$&$\mathbb{Z}_2 \oplus \mathbb{Z}_2$&$\mathbb{Z}_2 \oplus \mathbb{Z}_2 $&$\mathbb{Z}\oplus \mathbb{Z}$&0&0&0&$2\mathbb{Z}\oplus 2\mathbb{Z}$&0\\
                                    CII\qquad$R^{++}$&$R_6 \times R_6$&$0$& 0&$2\mathbb{Z}\oplus 2\mathbb{Z}$&0&$\mathbb{Z}_2 \oplus \mathbb{Z}_2$&$\mathbb{Z}_2 \oplus \mathbb{Z}_2$&$\mathbb{Z}\oplus \mathbb{Z}$&0\\
                                    DIII\qquad$R^{+-}$&$R_7$& 0&0&0&$2\mathbb{Z}$&0&$\mathbb{Z}_2$&$\mathbb{Z}_2$&$\mathbb{Z}$\\
                                    CI\qquad$R^{+-}$&$R_3$&0&$\mathbb{Z}_2$&$\mathbb{Z}_2$&$\mathbb{Z}$& 0&0&0&$2\mathbb{Z}$\\
                                    BDI\qquad$R^{+-}$&$R_1$&$\mathbb{Z}_2$&$Z$& 0&0&0&$2\mathbb{Z}$&0&$\mathbb{Z}_2$\\
                                    CII\qquad$R^{+-}$&$R_5$&0&$2\mathbb{Z}$&0&$\mathbb{Z}_2$&$\mathbb{Z}_2$&$\mathbb{Z}$& 0&0\\
                                    DIII \qquad $R^{-+}$&$R_3$&0&$\mathbb{Z}_2$&$\mathbb{Z}_2$&$\mathbb{Z}$& 0&0&0&$2\mathbb{Z}$\\
                                    CI\qquad$R^{-+}$&$R_7$& 0&0&0&$2\mathbb{Z}$&0&$\mathbb{Z}_2$&$\mathbb{Z}_2$&$\mathbb{Z}$\\
                                    BDI\qquad$R^{-+}$&$R_1 $&$\mathbb{Z}_2$&$\mathbb{Z}$& 0&0&0&$2\mathbb{Z}$&0&$\mathbb{Z}_2$\\
                                    CII\qquad$R^{-+}$&$R_5 $&0&$2\mathbb{Z}$&0&$\mathbb{Z}_2$&$\mathbb{Z}_2$&$\mathbb{Z}$& 0&0\\
                                    DIII,CI \qquad $R^{--}$&$C_0$&$\mathbb{Z}$& 0&$\mathbb{Z}$&0&$\mathbb{Z}$&0&$\mathbb{Z}$&0\\
                                    BDI\qquad$R^{--}$&$R_0 \times R_0$&$\mathbb{Z}\oplus \mathbb{Z}$&0&$0$& 0&$2\mathbb{Z}\oplus 2\mathbb{Z}$&0&$\mathbb{Z}_2 \oplus \mathbb{Z}_2$&$\mathbb{Z}_2 \oplus \mathbb{Z}_2$\\
                                    CII\qquad$R^{--}$&$R_4 \times R_4$&$2\mathbb{Z}\oplus 2\mathbb{Z}$&0&$\mathbb{Z}_2 \oplus \mathbb{Z}_2 $&$\mathbb{Z}_2 \oplus \mathbb{Z}_2 $&$\mathbb{Z}\oplus \mathbb{Z}$&0&0&0\\
                                    \hline\hline
                                  \end{tabular}
                                  \end{center}
                                \end{table*}

                                Next we consider the classification of the real classes. For the Hermitian Hamiltonian with multiple additional symmetries $\{ M_i \}$, where $\{M_i,H\}=0$ and $\{ M_i,M_j \}=2\delta _{i,j}$, following Ref.\cite{Furusaki}, we know conclusions for the classifying space for the Class AI and AII ($T$ only) or C and D ($C$ only):  Given that $M_iT=\eta _T TM_i$ or $M_iC=\eta _C CM_i$, we denote the relations as $M_i^{\eta _T}$ or $M_i^{\eta _C}$, and the number of $M_i^{+}$ and $M_i^{-}$ as $n^+$ and $n^-$. For class AI and AII, the classifying space $R_q$ shifts to $R_{q+n^+-n^-}$.  For class C and D, the classifying space $R_q$ shifts to $R_{q+n^--n^+}$.
                                Taking advantage of the above conclusions,  we can get the classifying space for the corresponding non-Hermitian systems with reflection symmetry.
                                In the following discussion, for convenience, we shall use $T$, $C$ and $S$ to represent $T_1$, $C_1$ and $S_1=T_1C_1$. Because $\Sigma$ ($\Sigma=\sigma_z\otimes 1$) commutes with both $T$ and $C$ in each class, we don't need to discuss their commutation relations separately in each class. For class AI, AII, C and D, $R^{\eta_T(\eta_C)}$ represent that $R_lT=\eta _T TR_l$ ($R_lC=\eta _C CR_l$).

                                    {\bf Class AI($R^-$):} Two symmetries $M$ and $\Sigma$ are added to the system of $H_a$. Since $T$ anticommutes with $\gamma_l$, $J$ and $R$, $M=JR\gamma_l$ anticommutes with $T$. It follows that $n^+=1$, $n^-=1$ and $\tilde{q}=q+n^+-n^-=0$. The classifying space is still $R_0$.

                                    {\bf Class AI($R^+$):} Since $T$ anticommutes with $\gamma_l$ and $J$, $M=JR\gamma_l$ commutes with $T$. Thus we have  $n^+=2$, $n^-=0$ and $\tilde{q}=q+n^+-n^-=2$. The classifying space $R_0$ shifts to $R_2$.

                                    {\bf Class AII($R^-$):} Since $T$ anticommutes with $\gamma_l$, $J$ and $R$, $M=JR\gamma_l$ anticommutes with $T$. Thus we have $n^+=1$, $n^-=1$ and $\tilde{q}=q+n^+-n^-=4$. The classifying space is still $R_4$.

                                    {\bf Class AII($R^+$):}  Since $T$ anticommutes with $\gamma_l$ and $J$, $M=JR\gamma_l$ commutes with $T$. Thus we have $n^+=2$, $n^-=0$ and $\tilde{q}=q+n^+-n^-=6$. The classifying space $R_4$ shifts to $R_6$.

                                    {\bf Class C($R^-$):} Since $C$ anticommutes with $R$ and $J$, $M=JR\gamma_l$ commutes with $C$. It follows that $n^+=2$, $n^-=0$ and $\tilde{q}=q+n^--n^+=4$. The classifying space $R_6$ shifts to $R_4$.

                                    {\bf Class C($R^+$):}  Since $C$ anticommutes with $J$, $M=JR\gamma_l$ anticommutes with $C$. Thus we have $n^+=1$, $n^-=1$ and $\tilde{q}=q+n^--n^+=6$. The classifying space is still $R_6$.

                                    {\bf Class D($R^-$):} Since $C$ anticommutes with $R$ and $J$, $M=JR\gamma_l$ commutes with $C$. It follows $n^+=2$, $n^-=0$ and $\tilde{q}=q+n^--n^+=0$. The classifying space $R_2$ shifts to $R_0$.

                                    {\bf Class D($R^+$):} Since $C$ anticommutes with $J$, $M=JR\gamma_l$ anticommutes with $C$. Thus we have $n^+=1$, $n^-=1$ and $\tilde{q}=q+n^--n^+=2$. The classifying space is still $R_2$.

                                 Once the classifying space is known, for example given by $R_q$, we can get that the system is characterized by $\pi_d(R_q)$. The classification results are shown in table III.  Similarly, we can analyze classes BDI, DIII, CI and CII with reflection symmetry and get the classifying space for these non-Hermitian systems (see the appendix B). The classification results are also listed in table III.

 \section{Examples }
      Now we discuss some examples of non-Hermitian topological phases protected by reflection symmetry. Particularly, we confine our study on 1D systems and give a complete list of all topologically nontrivial types in one dimension. We also construct explicitly topological invariants for our example Hamiltonians. For convenience, we use $R$ to represent $R_l$ in this section.

      {\bf 1. Class AIII ($R^-$).} For this class, the Hamiltonian should satisfy that
      $SH(k)=-H(k)S$ and $RH(-k)=H(k)R$, and we have $RS = - SR$.
      Consider a 2-band Hamiltonian having the following form:
      \begin{equation*}
         H(k)=\left[ \begin{array}{cc}
             0 & h(k)\\
             h(-k)& 0
             \end{array}
             \right ]  . \label{AIII}
         \end{equation*}
It is easy to check that the chiral and reflection operators are given by $S=\sigma_z$ and $R=\sigma_x$, respectively.  Taking
\[
h(k)= t_1e^{i\alpha}+t_2 e^{i\beta} e^{-ik},
\]
we get the non-Hermitian Hamiltonian
\begin{equation}
H(k) = ( t_1e^{i\alpha} + t_2 e^{i\beta} \cos k ) \sigma_x + t_2 e^{i\beta} \sin k \sigma_y.  \label{AIII-kmodel}
\end{equation}
From the table II, we know that the topological phase of the $\mathbb{Z}$-type system can be characterized by a winding number. Define the topological invariant
\begin{equation}
W=\frac{i}{2\pi}\int _0 ^{2\pi}\partial _k log(det(h(k))) .
\end{equation}
It is easy to check that $W=1$ for $\alpha=\frac{\pi}{3}$, $\beta=0$, $t_1=0.5$ and $t_2=1$,
              and $W=0$ for $\alpha=\frac{\pi}{3}$, $\beta=0$, $t_1=1.5$ and $t_2=1$, which indicates the system in different topological phases.

             {\bf 2. Class AI($R^+$).} According to the definition, the system has time-reversal and reflection symmetry.
             Consider a two-band Hamiltonian given by
              \begin{equation*}
                  H(k)=\left[ \begin{array}{cc}
                      h_1(k) & h_2(k)\\
                      h_2(-k)& h_1(-k)
                      \end{array}
                      \right ]  . \label{AI}
                  \end{equation*}
          with $h^*_{1,2}(k)=h_{1,2}(-k)$.
It is easy to check that the Hamiltonian fulfills $TH(-k)=H(k)T$ and $RH(-k)=H(k)R$, where $T=K$, $R=\sigma _x$ and $TR=RT$. We also have
                  \begin{equation*}
                      H(k)TR|\psi \rangle =TRH(k)|\psi \rangle =E^*(k)TR|\psi \rangle,
                  \end{equation*}
which   suggests that  the eigenvalues are either real or complex with
conjugate pairs. It follows that $detH(k)$ is real, and we can define the $\mathbb{Z}_2$ topological invariant as
                  \begin{equation}
                  D=sgn(det(H(k))).
                  \end{equation}
For the example Hamiltonian:
                  \begin{equation*}
                      H(k)=m\sigma_x+\alpha \sin k \sigma_y + i \beta  \sin k \sigma_z  + h \cos k \sigma_0,
                          \end{equation*}
we have
                          \begin{equation*}
                              det(H(k))=-m^2-(\alpha \sin k)^2+(\beta \sin k)^2+(h \cos k)^2.
                                  \end{equation*}
                               Consider the case with $\beta=h$.  If $\beta^2 > m^2 + \alpha^2$, we have $D=1$. If
                                  $\beta^2 < m^2 $, then $D=-1$. The $\mathbb{Z}_2$
                                  topological invariant $D=1$ or $D=-1$ characterizes topologically different phases.

          {\bf 3. Class D($R^+$).}  Consider the Hamiltonian given by
              \begin{equation*}
                  H(k)=\left[ \begin{array}{cc}
                      h_1(k) & h_2(k)\\
                      h_2(-k)& h_1(-k)
                      \end{array}
                      \right ]  . \label{DR+}
                  \end{equation*}
The Hamiltonian should fulfill the pseudo particle-hole and reflection symmetry, i.e.,  $CH(-k)=-H(k)C$ and $RH(-k)=H(k)R$. Taking $T=K$ and $R=\sigma _x$, the symmetries enforces  $h^*_{1,2}(k)= - h_{1,2}(-k)$. By observing
                  \begin{equation*}
                      H(k)CR|\psi \rangle=-CRH(k)|\psi \rangle =-E^*(k)CR|\psi \rangle,
                  \end{equation*}
we see the eigen energies appearing in pairs with the form of $a+ i b$ and $-a+ib$, where $a$ and $b$ are real. Since $detH(k)$ is real, similarly we can define $\mathbb{Z}_2$ topological invariant as $D=sgn(det(H(k)))$.
An example Hamiltonian is given by
      \begin{equation*}
      H(k)=i m \sigma_x + i \alpha \sin k \sigma_y + \beta  \sin k \sigma_z + i h \cos k\sigma_0 .
      \end{equation*}
It is straightforward to get
      \begin{equation*}
      det(H(k))=m^2+(\alpha \sin k)^2-(\beta \sin k)^2-(h \cos k)^2 .
      \end{equation*}
Consider the case with $\beta=h$.  If $\beta^2 > m^2 + \alpha^2$, we have $D=-1$. If
                                  $\beta^2 < m^2 $, then $D=1$.

          {\bf 4. Class BDI($R^{+-}$).} For all classes labeled by $R^{\eta_T \eta_C}$ which shall be discussed in the following text,  the Hamiltonian satisfies that $TH(-k)=H(k)T$, $CH(-k)=-H(k)C$, $SH(k)= - H(k)S$, and $RH(-k)=H(k)R$. Consider the Hamiltonian
              \begin{equation}
               H(k)=\left[ \begin{array}{cc}
                0 & h(k)\\
                h(-k)& 0
              \end{array}
              \right ]  . \label{h}
             \end{equation}
It is easy check that the chiral and reflection are fulfilled by taking $S=\sigma_z$ and $R=\sigma_x$.  Taking $T=K$ and $C=\sigma _z K$, we see that the T and C symmetries are fulfilled if $h^*(-k)=h(k)$. The additional constraint $RTH(k)=H(k)RT$ means that the two-band $H(k)$ is a Hermitian Hamiltonian, which suggests the two-band BDI($R^{+-}$) class must be a Hermitian system. A well known example is the SSH model described by
\begin{equation*}
                                  H(k)=(m+\alpha \cos k)\sigma_x+ \beta \sin k \sigma_y,
                                      \end{equation*}
for which the topological phases can be characterized by the winding number $W$.


  {\bf  5. Class BDI($R^{-+}$).}  For the Hamiltonian given by
                                      \begin{equation}
                                          H(k)=\left[ \begin{array}{cc}
                                              0 & h(k)\\
                                              -h(-k)& 0
                                              \end{array}
                                              \right ]   \label{h1}
                                          \end{equation}
                                      with $h^*(-k)=h(k)$, it is easy to check all symmetries are fulfilled with $T=K$, $C=\sigma _z K$, $S=\sigma_z$ and $R=\sigma_y$. We also have $\{R,T\}=0$ and $[R,C]=0$.  The topological phase can be characterized by
                                          $W=\frac{i}{2\pi}\int _0 ^{2\pi} \partial _k ln(det(h(k))$.
 For the example Hamiltonian
                                      \begin{equation}
                                      H(k)=i (m+\alpha \cos k)\sigma_y +i \beta \sin k \sigma_x, \label{R-+k}
                                      \end{equation}
                                       it follows that $W=-1$ when $\alpha=\beta=1$ and $m=0$, and $ W=1$ when $\alpha=-\beta=1$ and $m=0$.

  {\bf  6. Class CI($R^{+-}$).}
                               For the Hamiltonian given by
                               \begin{equation}
                                  H(k)=\left[ \begin{array}{cc}
                                      0 & h(k)\\
                                      -\sigma_y h^*(-k)\sigma_y & 0
                                      \end{array}
                                      \right ]  , \label{ci}
                                  \end{equation}
                               we have $T=\sigma _y \otimes  \sigma_y K$, $C=\sigma _x \otimes i \sigma_y K$ and $S=\sigma _z \otimes 1 $\cite{twogauge}. It is straightforward to check that the reflection operator is given by $R=\sigma_x \otimes \sigma _y$, fulfilling that $RH(k)=H(-k)R$,
                                  $RT=TR$ and $RC=-CR$. Then we get $RCH(k)=-H(k)RC$, which leads to $H^*(k)=-H(k)$, suggesting that $iH(k)$ is a  real matrix. According to the classification table, the topological phase should be characterized by a $\mathbb{Z}_2$ number, which can be defined as
                                  $D=sgn(det(ih(k)))$.
For the example Hamiltonian with $h(k)$ given by
                       \begin{equation*}
                           h(k)=mi\sigma_0+ i \beta  \sin k \sigma_x+\alpha \sigma_y+ i b \cos k  \sigma_z,
                       \end{equation*}
                                  we have
                                      \begin{equation*}
                                          det(ih(k))=m^2+\alpha^2-(\beta \sin k)^2-(b \cos k)^2.
                                      \end{equation*}
Considering the case with $\beta=b$, we have $D=1$ for $\beta^2 < m^2+\alpha^2$ and $D=-1$ for $ \beta^2 > m^2+\alpha^2 $.

                       {\bf 7. Class DIII($R^{-+}$).}
                       Consider the Hamiltonian  given by
                        \begin{equation}
                           H(k)=\left[ \begin{array}{cc}
                               0 & h(k)\\
                               \sigma _yh^*(-k)\sigma _y & 0
                               \end{array}
                               \right ]  . \label{DIII}
                           \end{equation}
                         It is not hard to find the T, C, and S symmetries are fulfilled by taking $T=\sigma _x \otimes i\sigma _y K$, $C=-\sigma _y \otimes \sigma _y K$ and  $S=\sigma _z \otimes 1 $\cite{twogauge}. Furthermore, we find that the reflection operator is given by $R=\sigma_x \otimes \sigma _y$, satisfying $RH(k)=H(-k)R$,
                           $RT=TR$ and $RC=-CR$. It follows $RTH(k)=H(k)RT$, which leads to  $H^*(k)=H(k)$, i.e., $H(k)$ is real matrix. According to the classification table, the topological phase is characterized by a $\mathbb{Z}_2$ topological invariant, which is defined by
                           $D=sgn(det(h(k))$.
For the example Hamiltonian
                           \begin{equation*}
                               h(k)=m\sigma_0+\beta  sink\sigma_x+ i \alpha  \sigma_y+b cosk \sigma_z,
                           \end{equation*}
we have
                               \begin{equation*}
                                   det(h(k))=m^2+\alpha^2-(\beta \sin k)^2-(b \cos k)^2 .
                               \end{equation*}
Considering the case with $\beta=b$, we have $D=-1$ for $\beta^2 > m^2 +\alpha^2$ and $D=1$ for $\beta^2 < m^2 +\alpha^2$.

                        {\bf 8. Class BDI($R^{++}$).}
                        For the Hamiltonian given by
                           \begin{equation}
                               H(k)=\left[ \begin{array}{cc}
                                   0 & h_1(k)\\
                                   h_2(k)& 0
                                   \end{array}
                                   \right ]   \label {h}
                               \end{equation}
                             with $h^*_{1,2}(-k)=h_{1,2}(k)$,  it is straightforward to check the T, C and S symmetries are fulfilled if we take $T=K$, $C=\sigma _z K$ and $S=\sigma _z $. Introducing the reflection operator as $R=\sigma_0 \otimes \sigma_x$,
we see that $RT=TR$ and $RC=CR$. The reflection symmetry requires $RH(k)=H(-k)R$, which leads to $\sigma_x h_{1,2}(-k)=h_{1,2}(k)\sigma_x$. It suggests that
                                $\sigma_x$ is a reflection operator for $h_{1,2}(k)$ and $K$ is the time reversal operator. Then both $h_{1}(k)$ and $h_{2}(k)$ belong to the class AI($R^+$), and each of them is characterized by a $\mathbb{Z}_2$ topological invariant. Correspondingly,
                                $H(k)$ is characterized by a $\mathbb{Z}_2 \oplus \mathbb{Z}_2$ topological invariant.
Consider the example Hamiltonian
                           \begin{equation*}
                               h_j(k)=m_j\sigma_x+ \alpha _j \sin k \sigma_y + i \beta _j\sin k \sigma_z + b_j \cos k \sigma_0,
                           \end{equation*}
                           we have
                               \begin{equation*}
                                   det(h_j(k))=(\beta_j \sin k)^2+(b_j \cos k )^2-m_j^2-(\alpha_j \sin k)^2.
                               \end{equation*}
                               Consider the case with $\beta_j=b_j$. If $\beta_j^2 > m_j^2 +\alpha_j^2$, we have $D_j=1$. If $\beta_j^2 < m_j^2$, then $D_j=-1$ ($j=1,2$). So the topological phase classified by a $\mathbb{Z}_2 \oplus \mathbb{Z}_2$ topological invariant.

                               {\bf 9. Class CII($R^{+-}$).} Considering the Hamiltonian given by  Eq.(\ref {h}) and taking $T=\sigma _0 \otimes i\sigma _yK$, $C=-\sigma _z \otimes i\sigma _y K$ and $S =\sigma _z \otimes 1$, we find that T, C and S symmetries are fulfilled if there exists an additional constraint:
                               \begin{equation}
                               \sigma_y h^*_{1,2}(k)=h_{1,2}(-k)\sigma_y. \label{T-CII}
                               \end{equation}
                               The above constraint condition suggests that the Hamiltonian $h_{1}(k)$ and $h_{2}(k)$ fulfill the time reversal symmetry separately. If we do not consider the reflection symmetry,
                               $h_{1,2}(k)$ belong to the class AII with
                               a $2\mathbb{Z}$ topological invariant \cite{Gong}, and consequently $H(k)$  is characterized by a
                               $2\mathbb{Z}\oplus 2\mathbb{Z}$ invariant.  Adding reflection symmetry with the reflection operator given by $R=\sigma_x \otimes 1$, we can check
                               $RT=TR$ and $RC=-CR$. The reflection symmetry $RH(k)=H(-k)R$ leads to $h_{1,2}(k)=h_{2,1}(-k)$, suggesting that only one of $h_{1,2}(k)$ is independent. So the topological phase of $H$ is characterized by a $2\mathbb{Z}$ topological invariant. For the example Hamiltonian
                               \begin{equation}
                                       H(k)=\left[ \begin{array}{cc}
                                           0 & h (k)\\
                                           h (-k)& 0
                                           \end{array}
                                           \right ]   \label{ciiR+-2}
                                       \end{equation}
                               with
                               \begin{equation}
                               h(k) = J_2 + i J_1e^{ik} \sigma_y+i\delta_1e^{ik}\sigma_x+i\delta_2e^{ik}\sigma_z, \label{ha(k)}
                               \end{equation}
                               we can define the topological invariant as
                               \begin{equation}
                               W=\frac{i}{2\pi}\int _{0}^{2\pi}dk\partial _k ln(det(h (k))). \label{W-CII}
                               \end{equation}
                               It is straightforward to get $W=-2$ for $0<J_2\ll J_1$ and $W=0$ for $0<J_1\ll J_2$ (we requires that $|\delta_1|,|\delta_2|\ll J_1,J_2 $ ).

                              {\bf 10. Class CII($R^{-+}$)}. Similar to the Class CII($R^{+-}$), for the Hamiltonian given by  Eq.(\ref {h}),
                               we have $T=\sigma _0 \otimes i\sigma _yK$, $C=-\sigma _z \otimes i\sigma _yK$
                                and $ S=\sigma _z \otimes 1$, and the Hamiltonian satisfies T, C and S symmetries if the constrain condition Eq.(\ref{T-CII}) is fulfilled. Introducing the reflection operator $R=\sigma_x \otimes \sigma_y$, we have
                               $RT=-TR$ and $RC=CR$. The reflection symmetry $RH(k)=H(-k)R$ leads to $\sigma_y h_{1,2}(k) \sigma_y=h_{2,1}(-k)$, and thus only one of $h_{1,2}(k)$ is independent. Similarly, the topological phase is also characterized by a $2\mathbb{Z}$ topological invariant. Consider the example Hamiltonian
                               \begin{equation}
                                H(k)=\left[ \begin{array}{cc}
                                    0 & h(k)\\
                                    \sigma _yh(-k)\sigma _y& 0
                                    \end{array}
                                    \right ]  . \label{CIIR-+}
                                \end{equation}
 with  $h(k)$ given by Eq.(\ref{ha(k)}).
                              The topological invariant is also given by Eq.(\ref{W-CII}). Consequently, we have $W=-2$ for $0<J_2\ll J_1$ and $W=0$ for $0<J_1\ll J_2$ (we requires that $|\delta_1|,|\delta_2|\ll J_1,J_2 $ ).

                                   \section{Bulk-edge correspondence}
                                       In the previous section, we construct model Hamiltonians in momentum space for all 1D non-trivial classes and give the definitions of corresponding topological numbers. In this section we discuss
                                       the bulk-edge correspondence by studying several examples of non-trivial classes. To study the bulk-edge correspondence, we need consider the model in the coordinate space under the open boundary condition (OBC).

                                       \subsection{Class AIII($R^-$) }
                                       Consider a lattice version of the model (\ref{AIII-kmodel}) described by the Hamiltonian:
                                       \begin{equation}
                                         \mathcal{H}=\sum_{n}[t_1 e^{i\alpha}a_n^{\dagger}b_n+t_1e^{i\alpha}b_n^{\dagger}a_n+t_2e^{i\beta}b_n^{\dagger}a_{n+1}+t_2e^{i\beta}a_{n+1}^{\dagger}b_{n}],
                                       \end{equation}
                                    which can be viewed as a non-Hermitian extension of the  SSH model \cite{SSH,Linhu2014}. For convenience, we set $t_1=t$ and $t_2=1$.  In the momentum space, the Hamiltonian transforms to
                                   \begin{equation}
                                   \begin{split}
                                    \mathcal{H}(k)&=(t e^{i\alpha}+ e^{i\beta}e^{-ik})a_k^{\dagger}b_k+(t e^{i\alpha}+ e^{i\beta}e^{ik})b_k^{\dagger}a_k\\
                                      &=(a_k^{\dagger},b_k^{\dagger})H(k)\left(
                                                                            \begin{array}{c}
                                                                              a_k \\
                                                                              b_k \\
                                                                            \end{array}
                                                                          \right) ,
                                    \end{split}   \label{a3}
                                   \end{equation}
                                   where $H(k)=h_{+}(k) \sigma_+ + h_{-}(k) \sigma_-$
                                   with $h_{+}(k) = t e^{i\alpha}+ e^{i\beta-ik}$ and $h_{-}(k)= t e^{i\alpha}+ e^{i\beta+ik}= h_{+}(-k)$, which is identical to  Eq.(\ref{AIII-kmodel}). The eigenvalue of the Hamiltonian is given by
                                   \[
                                   E=\pm \sqrt{h_+(k) h_+(-k)}= \pm \sqrt{t^2 e^{2 i\alpha}+ 2 t e^{i(\alpha+\beta)}\cos k + e^{2i\beta}}
                                   \]
                                   and the gap closes at $|t|=1$. The topological invariant is
                                   \begin{equation}
                                     \begin{split}
                                    W = \frac{i}{2\pi}\int _0^{2\pi} \text{d}k \partial _k ln(h_+) =  \frac{i}{2\pi}\int _0^{2\pi} \text{d}k\frac{\partial_kh_+}{h_+} .
                                     \end{split}
                                   \end{equation}

                                   \begin{figure}[h] 
                                    \includegraphics[width=3.3in]{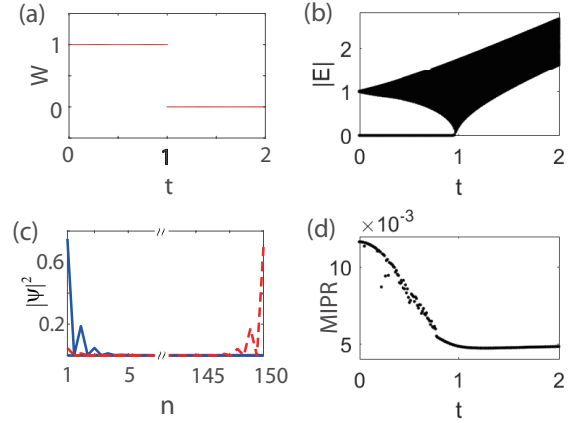}\\
                                    \caption{ (a) Phase diagram characterized by the winding number $W$ versus $t$. Here the model is under the periodic boundary condition with fixed parameters $\alpha=\pi/3$ and $\beta=0$. Figures (b), (c) and (d) are for the model with the same parameters but under the OBC with size $L=150$.  (b) The absolute values of eigenvalues versus $t$. At the region $t<1$, there exist a pair of zero modes. (c) The distribution of zero-mode wavefunctions of the system. Two zero modes are drawn in blue solid line and red dashed line, respectively. (d)  MIPR versus $t$, where MIPR is the average of IPRs of all eigenstates.} \label{fig1}.
                                   \end{figure}

                                    We show the topological invariant $W$ versus $t$ in Fig.{\ref{fig1}}a, which indicates a topological transition from the region of $t<1$ (with $W=1$) to of $t>1$ (with $W=0$).  In Fig{\ref{fig1}}b, we display the spectrum of the system under the OBC versus $t$. In the topologically non-trivial phase, we find that the system has double degenerate zero mode edge states as shown in  Fig{\ref{fig1}}c, whereas no zero mode edge state exists in the topologically trivial regime.
                                    It is clear that there is a bulk-edge correspondence in this class. In Fig{\ref{fig1}}d, we also display the mean inverse partition ration (MIPR), which is the average of inverse partition ratios (IPRs) of all eigenstates.  For a given state, its IPR is defined as
                                    \[
                                    IPR=\frac{\sum_{i}|\psi_i|^4}{(\sum_{i}|\psi _i|^2)^2},
                                    \]
                                    where $\psi_i$ is the wave function's amplitude at the site i. If the state is an extended state, its IPR shall tend to zero as the lattice size increases to the infinity limit. On the other hand, the IPR for a localized state remains to be finite even in the infinite size limit. We have checked that the increase of MIPR in the region $t<1$ is coming from the contribution of IPRs of zero mode edge states and no skin effect is found for other bulk states. Here we note that the non-Hermitian skin effect was dubbed to describe the emergence of skin-edge states of 1D non-reciprocal systems under the OBC, where all states are localized at one of edges \cite{WangZhong1}. The skin effect is usually induced by the non-reciprocal hopping processes and is sensitive to the boundary. Our results indicate that the skin edge states do not exit in the non-Hermitian systems with the reflection symmetry.
                                    \begin{figure}[h] 
                                    \includegraphics[width=3.3in]{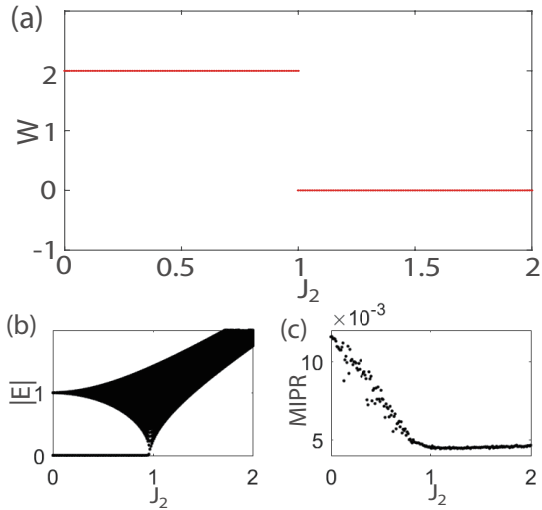}\\
                                    \caption{
                                    (a) Phase diagram characterized by the winding number $W$ versus $J_2$. Here the model is under the periodic boundary condition with fixed parameter $J_1=1$. Figures (b) and (c) are for the model with the same parameter but under the OBC with size $L=150$.  (b) The absolute values of eigenvalues versus $t$. At the region $J_2 <1$, there exist four-fold degenerate zero modes. (c)  MIPR versus $J_2$. } \label{fig2}.
                                   \end{figure}

                                   \subsection{Class CII($R^{+-}$) and Class CII($R^{-+}$) }
                                    Consider the lattice model described by the Hamiltonian:
                                    \begin{equation}
                                       \begin{split}
                                         \mathcal{H}=&\sum_{n,\sigma}J_2(a_{n,\sigma}^{\dagger}b_{n,\sigma}+b_{n,\sigma}^{\dagger}a_{n,\sigma})+\sum_{n}J_1(a_{n+1,\uparrow}^{\dagger}b_{n,\downarrow}-\\
                                         &a_{n+1,\downarrow}^{\dagger}b_{n,\uparrow}+b_{n,\uparrow}^{\dagger}a_{n+1,\downarrow}-b_{n,\downarrow}^{\dagger}a_{n+1,\uparrow}).
                                       \end{split}
                                       \end{equation}
                                   In the momentum space, the Hamiltonian can be represented as
                                   \begin{eqnarray*}
                                    \mathcal{H}(k) &=& J_2 \sum_{\sigma} ( a_{k,\sigma}^{\dagger} b_{k,\sigma}+b_{k,\sigma}^{\dagger} a_{k,\sigma}) \\
                                    & & + J_1 e^{-ik} ( a_{k,\uparrow}^{\dagger} b_{k,\downarrow} -  a_{k,\downarrow}^{\dagger} b_{k,\uparrow} ) \nonumber \\
                                    & & +  J_1 e^{ik}( b_{k,\downarrow}^{\dagger}a_{k, \uparrow} - b_{k,\downarrow}^{\dagger} a_{k,\uparrow} ).  \nonumber
                                    \end{eqnarray*}
                                    Alternatively, it can be also written as
                                    \begin{eqnarray}
                                      \mathcal{H}(k) = (a_{k,\uparrow}^{\dagger},a_{k,\downarrow}^{\dagger},b_{k,\uparrow}^{\dagger},b_{k,\downarrow}^{\dagger})H(k)\left(
                                                                            \begin{array}{c}
                                                                              a_{k,\uparrow}\\
                                                                              a_{k,\downarrow}\\
                                                                              b_{k,\uparrow}\\
                                                                              b_{k,\downarrow} \\
                                                                            \end{array}
                                                                          \right) ,  \label{a3}
                                   \end{eqnarray}
                                   where $H(k)$ is a $4 \times 4$ non-Hermitian matrix:
                                   \begin{equation}
                                     H(k)=\left(
                                                 \begin{array}{cc}
                                                   0 & h(k) \\
                                                   h(-k) & 0 \\
                                                 \end{array}
                                               \right)
                                               \end{equation}
                                   with
                                          \begin{equation} \label{aa}
                                              h(k)=\left(
                                            \begin{array}{cc}
                                              J_2 & J_1e^{-ik} \\
                                              -J_1e^{-ik} & J_2 \\
                                            \end{array}
                                          \right) .
                                   \end{equation}

                                   According to Eq.(\ref{ciiR+-2}), Eq.(\ref{ha(k)}) and Eq.(\ref{CIIR-+}) this Hamiltonian can represent class $CII(R^{+-})$ and class $CII(R^{-+})$ as terms of $\delta_1$ and $\delta_2$ are absent. And according to the classification table III, the topological invariant should be a $2\mathbb{Z}$ number, which only takes even numbers, and can be calculated by using Eq.(\ref{W-CII}). In Fig.{\ref{fig2}}a, we show the topological invariant $W$ versus $J_2$ by taking $J_1=1$.  A topological transition from the phase of $W=2$ to $W=0$ occurs at $J_2=1$.   The spectrum of the system under the OBC versus $J_2$ is displayed in Fig.{\ref{fig2}}b. We find that four-fold degenerate zero mode edge states exist in the topologically non-trivial phase corresponding to $W=2$, whereas no zero mode edge state exists in the topologically trivial regime. Our results indicate that a bulk-edge correspondence exists for this class. The MIPR shown in Fig.{\ref{fig2}}c indicates the absence of skin effect.

                                   \subsection{Class AI($R^+$)}
                                   Consider the lattice model
                                   \begin{equation}\label{ggh}
                                       \begin{split}
                                      \mathcal{H}=&\sum_{n}  \left [t ( a_n^{\dagger}b_n+ b_n^{\dagger}a_n) + t'(b_n^{\dagger}a_{n+1}+ a_{n+1}^{\dagger}b_{n})
                                      \right. \\
                                      & \left.  + t''(a^{\dagger}_n a_{n+1} + b^{\dagger}_{n+1}b_n) \right],
                                   \end{split}
                                   \end{equation}
                                   where the parameters $t$ ,$t'$ and $t''$ are real. When the non-Hermitian terms vanish, i.e., $t'' =0$, the model reduces to the well known SSH model.
                                   The Hamiltonian in momentum space is:
                                   \begin{eqnarray*}
                                     \mathcal{H}(k) = (a_k^{\dagger},b_k^{\dagger})H(k)\left(
                                                                             \begin{array}{c}
                                                                               a_k \\
                                                                               b_k \\
                                                                             \end{array}
                                                                           \right)
                                    \end{eqnarray*}
 with
                                    \begin{eqnarray}
                                      H(k)=\left(
                                               \begin{array}{cc}
                                                 t''e^{ik} & t+t'e^{-ik}\\
                                                 t+t'e^{ik} & t''e^{-ik} \\
                                               \end{array}
                                             \right)
                                   \end{eqnarray}
                                   or equivalently
                                   \[
                                   H(k)=t''\cos k\sigma_0+(t+t'\cos k)\sigma_x +t'\sin k\sigma_y +i t''\sin k\sigma_z.
                                   \]
                                    The topological invariant is
                                   \begin{equation*}
                                     D=\text{sgn(det($H(k)$))}.
                                   \end{equation*}

                                    \begin{figure}[h] 
                                    \includegraphics[width=3.3in]{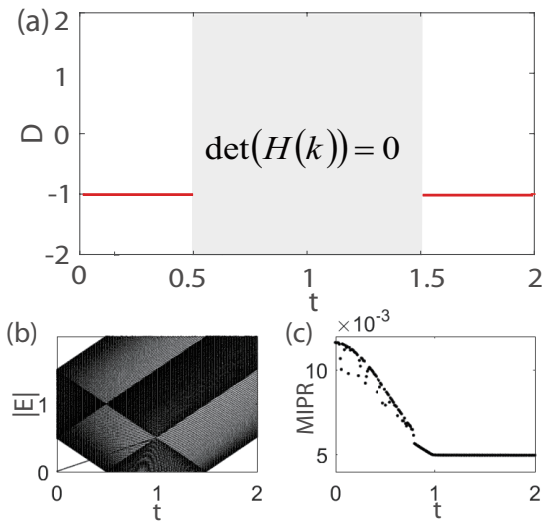}\\
                                    \caption{
                                    (a) Phase diagram characterized by the $\mathbb{Z}_2$ number $D$ versus $t$. Here the model is under the periodic boundary condition with fixed parameters $t'=1$ and $t''=0.5$. Figures (b) and (c) are for the model with the same parameters but under the OBC with size $L=50$.  (b) The absolute values of eigenvalues versus $t$. (c)  MIPR versus $t$.} \label{fig3}
                                   \end{figure}

                                  In Fig.{\ref{fig3}}a, we show the topological invariant $D$ versus $t$ by taking $t'=1$ and $t''=0.5$.  The spectrum of the system under the OBC versus $t$ is displayed in Fig.{\ref{fig3}}b.  In the topological non-trivial interval $t\in(0,0.5]$, the system has nonzero-mode edge states. In the topological non-trivial interval $t\in(1.5,2]$, the system doesn't have edge states, which can be read out from the distribution of MIPR shown in Fig.{\ref{fig3}}c. There is no bulk-edge correspondence  for the $\mathbb{Z}_2$ topological invariant $D$ in our studied model. The physical effect of the $D$ is not clear yet and it is possible to find a different topological number which may have the bulk-edge correspondence.

                                   \subsection{ Class BDI($R^{+-}$) and class BDI($R^{-+}$)}

                                    For class BDI($R^{+-}$), consider a 1D two-band Hamiltonian $H(k)$. It is easy to verify that $H(k)$ must be Hermitian (see section III). So it has a bulk-edge correspondence in this class.
                                    For class BDI($R^{-+}$), consider a 1D two-band Hamiltonian $H(k)$, e.g., given by Eq.(\ref{R-+k}). It is easy to verify that $H(k)$ must be anti-Hermitian because of Eq.(\ref{h1}). Since
                                    $H' = iH(k)$ is a Hermitian Hamiltonian with bulk-edge correspondence, so the anti-Hermitian Hamiltonian also has a bulk-edge correspondence in this class. Explicitly, we give a lattice model described by
                                    \begin{equation}
                                    H=\sum_{n} \left[ m(a_n^{\dagger}b_n- b_n^{\dagger}a_n) + \beta( b_n^{\dagger}a_{n+1}-a_{n+1}^{\dagger}b_n) \right],
                                    \end{equation}
                                    which corresponds to the model in the momentum space described Eq.(\ref{R-+k}) with $\alpha = -\beta$. It can be checked that this model under the OBC has pure imaginary spectrum except two zero edge modes at the topological regime $ |m| < |\beta|$.

                                     \section{Summary}

                                      In summary, we have studied non-Hermitian systems with the reflection symmetry and made a topological classification
                                     in the AZ classes in all dimensions.  In order to carry out the topological classification, we first mapped the non-Hermitian system into an enlarged Hermitian Hamiltonian, which has an enforced chiral symmetry, and thus the topological classification is equivalent to classifying Hermitian systems with both reflection symmetry and the enforced chiral symmetry. By determining the classifying space of the enlarged Hermitian Hamiltonian, we obtained the periodical classification tables of topological phases, which are summarized in table II and table III for the complex and real classes, respectively. Our results indicate the existence of  non-trivial phases in each dimension. Focusing on 1D systems, we constructed and analyzed concrete models for all non-trivial classes and defined the topological invariant for each non-trivial class.
                                     Then we studied the bulk-edge correspondence of some 1D non-Hermitian models. Our results indicate that there exits bulk-edge correspondence for all $\mathbb{Z}$ classes  characterized by winding numbers. However, we did not find  bulk-edge correspondence for the $\mathbb{Z}_2$ invariant D in our studied model of the class AI($R^+$). A future study may be the classification of non-Hermitian topological phases under crystal symmetries.

                                          \begin{acknowledgments}
C.-H. Liu  would thank C.-K. Chiu for very helpful discussions. The work is supported by NSFC under Grants No. 11425419, the National Key Research and Development Program of China (2016YFA0300600 and 2016YFA0302104) and the Strategic Priority Research Program (B) of the Chinese Academy of Sciences  (No. XDB07020000).
\end{acknowledgments}

                                            \section{appendix}
                                            \subsection{ Proof of equivalence between $H(k)$ and $U(k)$}

                                            Given that $H(k)$ satisfies the reflection symmetry,
                                            \begin{equation*}
                                                RH(k_1,k_2,...,-k_l,...,k_d)=H(k_1,k_2,...,k_l,...,k_d)R,
                                            \end{equation*}
                                            by using the polar decomposition $H(k)=U(k)P(k)$, with $P(k)$ being positive-definite Hermitian matrix,
                                            we get
                                            \begin{equation}
                                               \begin{split}
                                               &RU(k_1,k_2,...,-k_l,...,k_d)P(k_1,k_2,...,-k_l,...,k_d)=\\
                                               &U(k_1,k_2,...,k_l,...,k_d)P(k_1,k_2,...,k_l,...,k_d)R  . \label{ap1}
                                           \end{split}
                                            \end{equation}
                                            Then we get
                                            \begin{equation}
                                                \begin{split}
                                                &P^{\dagger}(k_1,k_2,...,-k_l,...,k_d)U^{\dagger}(k_1,k_2,...,-k_l,...,k_d)R^{\dagger}=\\
                                                &R^{\dagger}P^{\dagger}(k_1,k_2,...,k_l,...,k_d) U^{\dagger}(k_1,k_2,...,k_l,...,k_d) . \label{ap2}
                                            \end{split}
                                             \end{equation}
Left-multiplying Eq.(\ref {ap1}) by Eq.(\ref {ap2}) from both sides of equations and using $R^{\dagger}R =1$ and $U^{\dagger}U =1$, we get
                                       \begin{eqnarray*}
                                       &&P(k_1,k_2,...,-k_l,...,k_d)P(k_1,k_2,...,-k_l,...,k_d) \\
                                       &=&RP(k_1,k_2,...,k_l,...,k_d)P(k_1,k_2,...,k_l,...,k_d)R,
                                       \end{eqnarray*}
                                       where we have used $R^\dagger =R$ and $P^\dagger =P$. Then it follows
                                            \begin{equation}
                                       P(k_1,k_2,...,-k_l,...,k_d)=RP(k_1,k_2,...,k_l,...,k_d)R . \label{ap3}
                                            \end{equation}
                                            Together with Eq.(\ref {ap1}), we get that
                                            \begin{equation*}
                                               RU(k_1,k_2,...,-k_l,...,k_d)=
                                               U(k_1,k_2,...,k_l,...,k_d)R .
                                            \end{equation*}
                                             Defining a path
                                            \begin{equation*}
                                               \begin{split}
                                                   H_{\lambda}(k_1,k_2,...,-k_l,...,k_d)&=(1-\lambda) H +\lambda U\\
                                                   &=U((1-\lambda)P+\lambda),
                                               \end{split}
                                               \end{equation*}
                                            we have $H_0=H$ and $H_1=U$. For $0 \leq \lambda \leq 1$, $H_{\lambda}$ is reversible on the path
                                            and satisfies the reflection symmetry.

                                            \subsection{ Classification of non-Hermitian systems for classes BDI, DIII, CII and CI with reflection symmetry}

                                             First we consider the Hermitian Hamiltonian of Classes BDI, DIII, CII and CI (both $T$ and $C$) with multiple additional symmetries $\{ M_i \}$, where $\{M_i,H\}=0$ and $\{ M_i,M_j \}=2\delta _{i,j}$.
                                             Given that $M_iT=\eta _T TM_i$ and $M_iC=\eta _C CM_i$, we denote the relations as $M_i^{\eta _T\eta _C}$ and the number of $M_i^{\eta _T\eta _C}$ as $n^{\eta _T\eta _C}$. Following Ref.\cite{Furusaki}, we see that the classifying space $R_q$ shifts to classifying space represented by extension problem $Cl_{0,\tilde{q}}\otimes Cl_{0,m}\rightarrow Cl_{0,\tilde{q}+1}\otimes Cl_{0,m}$ with $\tilde{q}=q+n^{+-}-n^{-+}$ and
                                \begin{equation}
                                    m=\left\{
                                                 \begin{array}{lr}
                                                 n^{--}-n^{++} &  \text{(DIII and CI)}  \\
                                                 n^{++}-n^{--} &  \text{(BDI and CII)}
                                                 \end{array}
                                    \right. .
                                    \end{equation}
                                    The classifying space of the extension problem is listed in table IV.

                                    \begin{table}[h]
                                        \caption{\label{tab:table2}The classifying space of class BDI, DIII, CI and CII with all TRS, PHS and reflection symmetries. And it is a function of ($m,\tilde{q}$).}
                                      \begin{ruledtabular}
                                      \begin{tabular}{ccc}
                                        m(mod 8)&Classifying Space&$\pi_0$\\
                                        \hline
                                        0&$R_{\tilde{q}}$&$\pi_0$($R_{\tilde{q}}$)\\
                                        1&$R_{\tilde{q}}\times R_{\tilde{q}}$&$\pi_0$($R_{\tilde{q}}$)$\oplus\pi_0$($R_{\tilde{q}}$)\\
                                        2&$R_{\tilde{q}}$&$\pi_0$($R_{\tilde{q}}$)\\
                                        3&$C_{\tilde{q}}$&$\pi_0$($C_{\tilde{q}}$)\\
                                        4&$R_{\tilde{q}+4}$&$\pi_0$($R_{\tilde{q}+4}$)\\
                                        5&$R_{\tilde{q}+4}\times R_{\tilde{q}+4}$&$\pi_0$($R_{\tilde{q}+4}$)$\oplus\pi_0$($R_{\tilde{q}+4}$)\\
                                        6&$R_{\tilde{q}+4}$&$\pi_0$($R_{\tilde{q}+4}$)\\
                                        7&$R_{\tilde{q}+4}$&$\pi_0$($R_{\tilde{q}+4}$)\\
                                        \end{tabular}
                                      \end{ruledtabular}
                                    \end{table}

                                   With the help of the above conclusions,  we can get the classifying space for the corresponding non-Hermitian systems with reflection symmetry.
                                   For classes BDI, DIII, CI and CII, $R^{\eta_T\eta_C}$ represent that $R_lT=\eta _T TR_l$ and $R_lC=\eta _C CR_l$.

                                    {\bf Class DIII($R^{--}$):} $M=JR\gamma_l$ anticommutes with $T$ and commutes with $C$. $n^{++}=1$, $n^{-+}=1$ and  $n^{+-}=n^{--}=0$, then $m=n^{--}-n^{++}=-1$ and $\tilde{q}=q+n^{+-}-n^{-+}=2$. The classifying space $R_3$ shifts to $C_0$ according to table IV.

                                    {\bf Class DIII($R^{-+}$):} $M=JR\gamma_l$ anticommutes with $T$ and $C$. $n^{++}=1$, $n^{--}=1$ and  $n^{+-}=n^{-+}=0$, then $m=n^{--}-n^{++}=0$ and $\tilde{q}=q+n^{+-}-n^{-+}=7$. The classifying space is still $R_3$ according to table IV.

                                    {\bf Class DIII($R^{+-}$):} $M=JR\gamma_l$ commutes with $T$ and $C$. $n^{++}=2$, $n^{-+}=n^{+-}=n^{--}=0$, then $m=n^{--}-n^{++}=-2$ and $\tilde{q}=q+n^{+-}-n^{-+}=3$. The classifying space $R_3$ shifts to $R_7$ according to table IV.

                                    {\bf Class DIII($R^{++}$):} $M=JR\gamma_l$ commutes with $T$ and anticommutes with $C$. $n^{++}=1$, $n^{+-}=1$ and $n^{-+}=n^{--}=0$, then $m=n^{--}-n^{++}=-1$ and $\tilde{q}=q+n^{+-}-n^{-+}=4$. The classifying space $R_3$ shifts to $C_0$ according to table IV.

                                    {\bf Class CI($R^{--}$):} $M=JR\gamma_l$ anticommutes with $T$ and commute with $C$. $n^{++}=1$, $n^{-+}=1$ and $n^{+-}=n^{--}=0$, then $m=n^{--}-n^{++}=-1$ and $\tilde{q}=q+n^{+-}-n^{-+}=6$. The classifying space $R_7$ shifts to $C_0$ according to table IV.

                                    {\bf Class CI($R^{-+}$):} $M=JR\gamma_l$ anticommutes with $T$ and $C$. $n^{++}=1$, $n^{--}=1$ and $n^{+-}=n^{-+}=0$, then $m=n^{--}-n^{++}=0$ and $\tilde{q}=q+n^{+-}-n^{-+}=7$. The classifying space is still $R_7$ according to table IV.

                                    {\bf Class CI($R^{+-}$):} $M=JR\gamma_l$ commutes with $T$ and $C$. $n^{++}=2$ and  $n^{+-}=n^{-+}=n^{--}=0$, then $m=n^{--}-n^{++}=-2$ and $\tilde{q}=q+n^{+-}-n^{-+}=7$. The classifying space $R_7$ shifts to $R_{11} \simeq R_3$ according to table IV.

                                    {\bf Class CI($R^{++}$):} $M=JR\gamma_l$ commutes with $T$ and anticommutes with $C$. $n^{++}=1$, $n^{+-}=1$ and $n^{-+}=n^{--}=0$, then $m=n^{--}-n^{++}=-1$ and $\tilde{q}=q+n^{+-}-n^{-+}=8$. The classifying space $R_7$ shifts to $C_8 \simeq C_0$ according to table IV.

                                    {\bf Class BDI($R^{--}$):} $M=JR\gamma_l$ anticommutes with $T$ and commutes with $C$. $n^{++}=1$, $n^{-+}=1$ and $n^{+-}=n^{--}=0$, then $m=n^{++}-n^{--}=1$ and $\tilde{q}=q+n^{+-}-n^{-+}=0$. The classifying space $R_1$ shifts to $R_0\times R_0$ according to table IV.

                                    {\bf Class BDI($R^{-+}$):} $M=JR\gamma_l$ anticommutes with $T$ and $C$. $n^{++}=1$, $n^{--}=1$ and $n^{+-}=n^{-+}=0$, then $m=n^{++}-n^{--}=0$ and $\tilde{q}=q+n^{+-}-n^{-+}=1$. The classifying space is still $R_1$ according to table IV.

                                    {\bf Class BDI($R^{+-}$):} $M=JR\gamma_l$ commutes with $T$ and $C$. $n^{++}=2$, $n^{-+}=n^{+-}=n^{--}=0$, then $m=n^{++}-n^{--}=2$ and $\tilde{q}=q+n^{+-}-n^{-+}=0$. The classifying space is still $R_1$ according to table IV.

                                    {\bf Class BDI($R^{++}$):} $M=JR\gamma_l$ commutes with $T$ and anticommutes with $C$. $n^{++}=1$, $n^{+-}=1$ and $n^{-+}=n^{--}=0$, then $m=n^{++}-n^{--}=1$ and $\tilde{q}=q+n^{+-}-n^{-+}=2$. The classifying space $R_1$ shifts to $R_2\otimes R_2$ according to table IV.

                                    {\bf Class CII($R^{--}$):} $M=JR\gamma_l$ anticommutes with $T$ and commutes with $C$. $n^{++}=1$, $n^{-+}=1$ and  $n^{+-}=n^{--}=0$, then $m=n^{++}-n^{--}=1$ and $\tilde{q}=q+n^{+-}-n^{-+}=4$. The classifying space $R_5$ shifts to $R_4\times R_4$ according to table IV.

                                    {\bf Class CII($R^{-+}$):} $M=JR\gamma_l$ anticommutes with $T$ and $C$. $n^{++}=1$, $n^{--}=1$ and  $n^{+-}=n^{-+}=0$, then $m=n^{++}-n^{--}=0$ and $\tilde{q}=q+n^{+-}-n^{-+}=5$. The classifying space is still $R_5$ according to table IV.

                                    {\bf Class CII($R^{+-}$):} $M=JR\gamma_l$ commutes with $T$ and $C$. $n^{++}=2$, $n^{-+}=n^{+-}=n^{--}=0$, then $m=n^{++}-n^{--}=2$ and $\tilde{q}=q+n^{+-}-n^{-+}=5$. The classifying space is still $R_5$ according to table IV.

                                    {\bf Class CII($R^{++}$):} $M=JR\gamma_l$ commutes with $T$ and anticommute with $C$. $n^{++}=1$, $n^{+-}=1$ and  $n^{-+}=n^{--}=0$, then $m=n^{++}-n^{--}=1$ and $\tilde{q}=q+n^{+-}-n^{-+}=6$. The classifying space $R_5$ shifts to $R_6\times R_6$ according to table IV.


\begin{thebibliography}{99}
\bibitem{twogauge} We note that T and C have $U(1)$ gauge freedoms which do not change values of $T^2$ and $C^2$ and their commutation relations with the Hamiltonian. Nevertheless, T and C  have different commutation relations if we choose different gauges. When we choose the gauge to satisfy $S^2=1$, we denote it as the AZ gauge. In the AZ gauge, if we have  $TC=\eta CT$ $(\eta =\pm 1)$, then $S^2=(TC)^2=\eta T^2C^2=1$. When $T^2$ and $C^2$ have different signs, T should anticommute with C. On the other hand, for the convenience of topological classification, one usually chooses another gauge to enforce $[T,C]=0$.  By choosing such a gauge, we always have $[T,C]=0$, however $S^2$ may take $1$ or $-1$ depending on the value of $T^2$ and $C^2$.
\end{thebibliography}
\end{document}